\documentclass[runningheads]{llncs}

\usepackage[T1]{fontenc}
\usepackage{dirtree}

\usepackage{cite}
\usepackage{amsmath,amssymb,amsfonts}
\usepackage{float}
\usepackage{hyperref}
\usepackage{algorithmic}
\usepackage{textcomp}
\usepackage{xcolor}
\usepackage{verbatim}
\usepackage{listings}
\usepackage{xcolor,colortbl}
\usepackage{todonotes}
\usepackage{tcolorbox}
\usepackage{csquotes}
\usepackage{ifthen}
\usepackage{caption}
\usepackage{subcaption}
\usepackage[ruled, lined, titlenotnumbered]{algorithm2e}
\usepackage{orcidlink}

\usepackage{listings}
\usepackage{fontawesome}

\newcommand{\TwoSymbolsAndText}[3]{%
  #1\quad #2 \quad #3%
}

\usepackage[firstpage]{draftwatermark} 
\usepackage{graphicx}

\tcbuselibrary{listings,skins}
\definecolor{White}{rgb}{1,1,1}
\definecolor{codegreen}{rgb}{0,0.6,0}
\definecolor{codegray}{rgb}{0.5,0.5,0.5}
\definecolor{codepurple}{rgb}{0.58,0,0.82}
\definecolor{backcolour}{rgb}{0.95,0.95,0.92}
\definecolor{pblue}{rgb}{0.13,0.13,1}
\definecolor{lightblue}{rgb}{0.13,0.13,0.6}
\definecolor{pgreen}{rgb}{0,0.5,0}
\definecolor{pred}{rgb}{0.9,0,0}
\definecolor{pgrey}{rgb}{0.46,0.45,0.48}
\definecolor{mediumslateblue}{rgb}{0.48, 0.41, 0.93}
\definecolor{electricviolet}{rgb}{0.56, 0.0, 1.0}
\definecolor{LightCyan}{rgb}{0.88,1,1}
\definecolor{DarkBlue}{rgb}{0,0,0.2}
\definecolor{lightgreen}{rgb}{0.5,1,0.5}

\lstdefinestyle{mystyle}{
  backgroundcolor=\color{backcolour},   
  commentstyle=\color{codegreen},
  keywordstyle=\color{magenta},
  numberstyle=\tiny\color{black},
  stringstyle=\color{codepurple},
  basicstyle=\ttfamily\footnotesize,
  breakatwhitespace=false,     
  breaklines=true,         
  captionpos=b,          
  keepspaces=true,         
  numbers=left,          
  numbersep=5pt,          
  showspaces=false,        
  showstringspaces=false,
  showtabs=false,          
  tabsize=2
}

\lstdefinestyle{myCustomIvyStyle}{
  aboveskip=0em,
  belowskip=0em,
  numbers=left,
  stepnumber=1,
  numbersep=20pt,
  tabsize=2,
  mathescape,
  basicstyle=\scriptsize,
  showspaces=false,
  showstringspaces=false,
  keywordstyle=\color{pblue},
  keywordstyle=[2]{\color{mediumslateblue}},
  keywordstyle=[3]{\color{electricviolet}},
  identifierstyle=\color{black},
  commentstyle=\itshape\color{pgreen},
  stringstyle=\color{pred},
  morekeywords={include,then,after,type,String,action,object,variant,of,import, instance, around, if, while, implement, else, before,ensure, module, returns, return,python, implementation},
  morekeywords=[2]{require,var,call, destructor,instantiate,struct, , execute,is_set,value,set,ip,ipv6,  interpret, virtual,relation, seconds, microseconds, milliseconds,function},
  morekeywords=[3]{this, stream_pos,stream_data,trans_params_struct,cid,quic_packet_type,bool,frame,stream_id,quic_packet,endpoint, microsecs, pkt_num},
  extendedchars=true,
  morecomment=[l]{\#}
}

\lstdefinestyle{myCustomIvyStyleHardcoded}{
  aboveskip=0em,
  belowskip=0em,
  firstnumber=10,
  numbers=left,
  stepnumber=1,
  numbersep=20pt,
  tabsize=2,
  basicstyle=\scriptsize,
  showspaces=false,
  showstringspaces=false,
  keywordstyle=\color{pblue},
  keywordstyle=[2]{\color{mediumslateblue}},
  keywordstyle=[3]{\color{electricviolet}},
  identifierstyle=\color{black},
  commentstyle=\itshape\color{pgreen},
  stringstyle=\color{pred},
  morekeywords={include,then,after,type,String,action,object,variant,of,import, instance, around, if, while, implement, else, before,ensure, module, returns, return,python, implementation},
  morekeywords=[2]{require,var,call, destructor,instantiate,struct, , execute,is_set,value,set,ip,ipv6,  interpret, virtual,relation, seconds, microseconds, milliseconds},
  morekeywords=[3]{this, stream_pos,stream_data,trans_params_struct,cid,quic_packet_type,bool,frame,stream_id,quic_packet,endpoint, microsecs, pkt_num},
  extendedchars=true,
  morecomment=[l]{\#}
}

\newtcblisting[use counter=inputPrg]{codeInput}[3]{
  listing only,
  listing style=myCustomIvyStyle,
  size=title,
  arc=1mm,
  enhanced jigsaw,
  coltitle=White,
  boxrule=0.5mm,
  title=\TwoSymbolsAndText{\faCode}{%
  \ifthenelse{\equal{#2}{}}{}{\textbf{\thetcbcounter. #2}}%
  }{\faCode},
  label=inputPrg:#3
}

\newtcblisting[use counter=inputPrg]{codeInputHard}[3]{
  listing only,
  listing style=myCustomIvyStyleHardcoded,
  size=title,
  arc=1mm,
  enhanced jigsaw,
  coltitle=White,
  boxrule=0.5mm,
  label=inputPrg:#3
}

\usepackage{tikz}

\begin{document}
\SetWatermarkText{\hspace*{3.7in}\raisebox{8.55in}{\href{https://eapls.org/pages/artifact_badges/}{\mbox{\includegraphics[width=1.4cm]{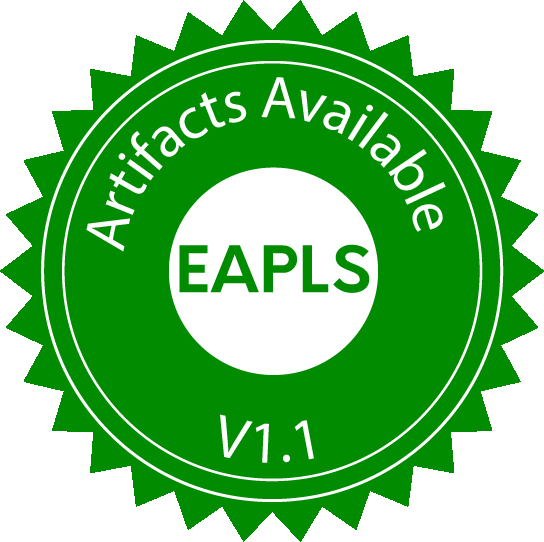}\includegraphics[width=1.4cm]{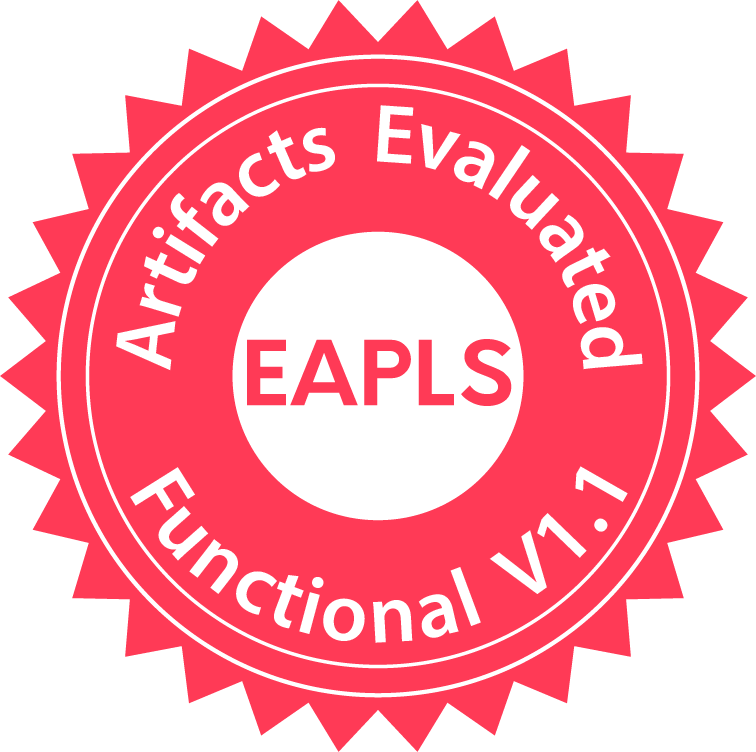}}}}}
\SetWatermarkAngle{0}

\title{Network Simulator-centric Compositional Testing}

\author{Tom Rousseaux \orcidlink{0009-0004-9590-3241} \thanks{carried out the research/implementation about network simulators} \and
Christophe Crochet \orcidlink{0000-0001-8635-2098}\thanks{carried out the research/implementation about formal specification of time properties \& implementation in PFV \& carried out the experiments on \texttt{QUIC}} \and 
John Aoga \orcidlink{0000-0002-7213-146X} \and 
Axel Legay \orcidlink{0000-0003-2287-8925} }

\authorrunning{T. Rousseaux et al.}

\institute{
INGI, ICTEAM, Université catholique de Louvain, Place Sainte Barbe 2, L05.02.01, 1348 Louvain-La-Neuve, Belgium\\
\email{\{firstname.lastname\}@uclouvain.be}}
\maketitle        

\begin{abstract}
    \sloppy{This article introduces a novel methodology, Network Simulator-centric Compositional Testing (NSCT), to enhance the verification of network protocols with a particular focus on time-varying network properties. NSCT follows a Model-Based Testing (MBT) approach. These approaches usually struggle to test and represent time-varying network properties. NSCT also aims to achieve more accurate and reproducible protocol testing. It is implemented using the Ivy tool and the Shadow network simulator. This enables online debugging of real protocol implementations. A case study on an implementation of \texttt{QUIC} (\textit{picoquic}) is presented, revealing an error in its compliance with a time-varying specification. This error has subsequently been rectified, highlighting NSCT's effectiveness in uncovering and addressing real-world protocol implementation issues. The article underscores NSCT's potential in advancing protocol testing methodologies, offering a notable contribution to the field of network protocol verification.}
    \keywords{Model-Based Testing, Time-varying Network Properties, Software verification and validation, Formal Specifications, Network Simulator, Internet protocols, \texttt{QUIC}, Concrete Implementation, Adverse Stimuli}
\end{abstract}

\section{Introduction}\label{sec:intro}

Ensuring the safety and effectiveness of systems is paramount. One way to achieve this goal is through the use of model-checking approaches. These approaches employ mathematical models of the system and exhaustively check the specifications against all possible behaviors of the system. Examples of such approaches include \textit{SPIN} \cite{spin97, lounas2017formal} and \textit{NUSMV}\cite{nusmv93,classen2011symbolic}, which use Linear-Temporal Logic (LTL) \cite{DBLP:conf/focs/Pnueli77} or Computation Tree Logic (CTL) \cite{clarke1981design} to describe specifications. First, model checking results were applied to mathematical models of the system under validation. However, over the last decades, we have seen the emergence of techniques applied directly to implementations \cite{DBLP:books/daglib/0007403-2}. An inherent hurdle in model-checking lies in the state-space explosion dilemma triggered by exhaustive exploration of the entire state-space.

To tackle this challenge, researchers have proposed Statistical Model Checking (LLTYSG19, LL20). This approach entails simulating the system and using statistical algorithms to ascertain whether it meets a measurable specification within a finite execution with a certain probability and given confidence. The approach, which has been implemented in tools such as \textit{UPPAAL-SMC}\cite{david2015uppaal,larsen2009uppaal,10.1145/2933575.2934574, 7911867} and \textit{PLASMA}\cite{plasma13, legay2014statistical,7911872}, has been applied on a wide range of case studies \cite{LL20}. Statistical Model Checking is primarily employed for validating properties on mathematical models. Except for specific instances like in \cite{DBLP:conf/cav/NgoLJ16}, direct validation on code has been rarely proposed. Generating fair traces from the code required by the statistical algorithm is challenging.
Furthermore, with few exceptions as seen in \cite{CDL10}, specifications are typically formulated using Bounded LTL. This representation is inadequate for describing the specifications of a network protocol. In this paper, we propose an approach that leverages the principle of simulation akin to Statistical Model Checking, but within the framework of specifications described in a sophisticated language and directly validated against the implementation. 
This approach, known as model-based testing \cite{offutt1999generating}, offers a more scalable solution. Our approach also introduces specific techniques to address the challenges involved in protocol verification.

In protocol verification, traditional methodologies rely on multiple independent implementations and interoperability testing to validate protocol designs. However, comprehensive model-based verification is often lacking in well-known approaches. Ivy, a notable exception, allows working with protocol implementations and adversarial stimuli.

Ivy's mathematical model\cite{Padon_McMillan_Panda_Sagiv_Shoham_2016} serves as the language describing the system specifications, whereas model-checking approaches typically involve two mathematical models, one for the system and one for the specifications.

Network-centric Compositional Testing (NCT) \cite{McMillan_Zuck_2019} is an emerging methodology that was introduced within Ivy. NCT introduces a formal statement of a protocol standard, allowing effective testing of implementations for compliance, not just interoperability. NCT uses formal specifications of protocols to automatically create testing tools. These tools generate random test cases by solving constraints with the help of an SMT solver. This enables adversarial testing in real-world environments, uncovering compliance issues and ambiguities in standard protocol documents (RFCs). NCT uncovered errors and vulnerabilities in the real-world protocol \texttt{QUIC}\cite{9000}, proving its effectiveness.

Although NCT serves as a foundation for network-centric protocol verification, it does not address time-varying network properties. Time-varying network properties describe the timed aspects of network protocols. These properties include internal timeouts to, for example, trigger a packet retransmission. These are involved in properties that are more complex to model, such as congestion control schemes in retransmission mechanisms.

Ivy deterministically generates output packets from input packets, but does not provide computation time requirements.
This duration can non-deterministi-\\cally exceed (or not) protocol timeouts.
This can impact the inputs, for example by triggering a packet retransmission.
This non-determinism prevents the reproducibility of the experiments.
If Ivy discovers a bug in an implementation, Ivy is not necessarily able to reproduce it.

We have developed a new approach called \textit{Network Simulator-centric Compositional Testing} (NSCT) to address these limitations. NSCT is designed to focus on verifying time-varying network properties in network protocols and ensuring experimental reproducibility. 

Our method extends Ivy to support time-related features, providing a more network-centric approach. Additionally, we have integrated the Shadow network simulator which allows online debugging of protocol implementations. This integration ensures the determinism and reproducibility of the experiments. We have successfully applied our approach to test the \texttt{QUIC} protocol and have demonstrated its effectiveness in verifying time-varying network properties. NSCT reveals an error on the \textit{picoquic} implementation and the \texttt{QUIC} idle timeout connection termination. This approach brings advancements to the network community by enabling more detailed and accurate protocol testing and verification. It enables specifying loss detection and congestion control in \texttt{QUIC} defined by \texttt{RFC9002} \cite{9002}.

The remainder of this paper is structured as follows. Section \ref{sec:back} provides background information on different verification types and the emergence of Ivy. Section \ref{sec:approach} outlines our methodology, providing details about \textit{Network Simulator-centric Compositional Testing} (NSCT).  Section \ref{sec:quicverif} delves into the verification of time-varying network properties in \texttt{QUIC}. Then, Section \ref{sec:futur} discusses our findings and proposes avenues for future work. Finally, Section \ref{sec:conclo} leads to the related work and the conclusion.

\section{Background}\label{sec:back}

There are two main ways to create adversarial tests for network protocols: with\cite{bozic2018formal,paris2009automatic,veanes2008model} or without\cite{lee2015gatling,cadar2008klee,rath2018interoperability,bhargavan2013implementing,chudnov2018continuous} checking compliance with a standard. Approaches that do not check compliance with a standard include fuzz testing\cite{lee2015gatling}, white-box testing\cite{cadar2008klee, rath2018interoperability}, and other methods that create a verified reference implementation\cite{bhargavan2013implementing} or prove properties of an existing implementation\cite{chudnov2018continuous}. On the contrary, approaches that verify compliance with a standard, also known as Model-Based Testing (MBT)\cite{offutt1999generating}, involve constructing an abstract model with Finite State Machines (FSMs) to explore and generate test scenarios\cite{bozic2018formal,paris2009automatic,veanes2008model}. However, the incorporation of data into FSMs adds significant complexity and challenges to these formalisms \cite{McMillan_Zuck_2019}.

Network-centric compositional (NCT) approaches avoid the use of FSMs. To grasp this and our approach, which builds upon NCT, it is crucial to understand the functioning of the \textit{Ivy tool} that implements them.  

\paragraph{\textbf{Ivy}} is a verification tool implementing multiple proving techniques \cite{Padon_McMillan_Panda_Sagiv_Shoham_2016,McMillan_Padon_2020}.  It is used to correct the design and implementation of algorithms and distributed protocols. It supports modular specifications and implementation. Ivy is used to interactively verify the safety properties of infinite-state systems. Ivy introduced a Relational Modeling Language (RML). This language allows describing the state of a program using formulas from first-order logic and uses relations (boolean predicate), functions, modules, and type objects as the main abstractions to represent the state of the system. Let us illustrate the functioning \textit{Ivy} using a running protocol example. 

\paragraph{\textbf{\texttt{MiniP}  (Minimalist Protocol)}} is a simple protocol. \texttt{MiniP} defines packets that contain frames. Any packet must contain exactly two frames. Three types of frames are defined: \texttt{PING}, \texttt{PONG}, and \texttt{TIMESTAMP} frames. \texttt{PING} frame contains a four-byte string representing the word "ping". \texttt{PONG} frame also contains a four-byte string expressing the word "pong". The \texttt{PING} frame or the \texttt{PONG} frame must be present in a packet. Finally, the \texttt{TIMESTAMP} frame contains an eight-byte unsigned integer representing the moment, in milliseconds, when the packet is sent. This frame must be present in all packets.

\begin{figure}
   \centering
   \begin{subfigure}[b]{0.49\textwidth}
     \centering
     \includegraphics[width=1\textwidth]{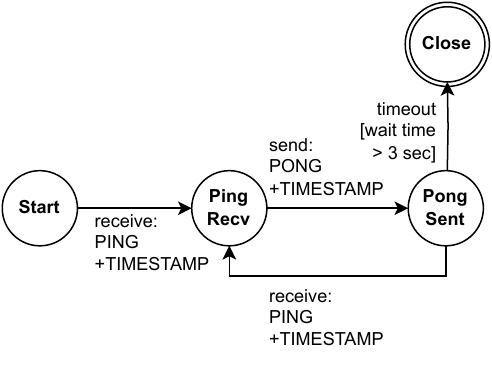}
     \caption{\texttt{MiniP} server FSM}
     \label{fig:y equals x}
   \end{subfigure}
   \hfill
   \begin{subfigure}[b]{0.49\textwidth}
     \centering
     \includegraphics[width=\textwidth]{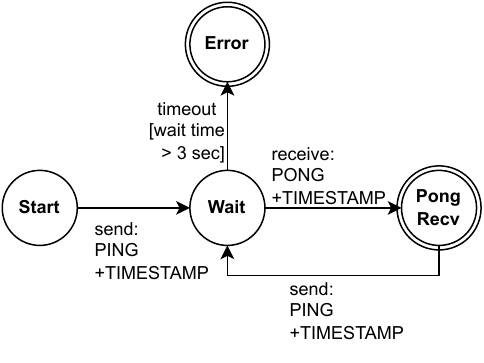}
     \caption{\texttt{MiniP} client FSM}
     \label{fig:five over x}
   \end{subfigure}
    \caption{\texttt{MiniP} Finite state machines (FSM)}
    \label{fig:fsm_minip}
    \vspace*{-1cm}
\end{figure}

Figure \ref{fig:fsm_minip} represents the finite state machines (FSM) of \texttt{MiniP}. The client starts by sending a packet containing the \texttt{PING} frame followed by the \texttt{TIMESTAMP} frame as payload. The server must then respond within three seconds, with a packet containing the \texttt{PONG} frame followed by the \texttt{TIMESTAMP} frame. This exchange continues until the client stops the connection. The client terminates the connection by not transmitting any packets for more than three seconds.

\paragraph{\textbf{Some MiniP components implementation with Ivy}} The first and most important components to implement for this protocol are frames. In Ivy, a frame is implemented using the \textit{type object}. Listing \ref{inputPrg:state} provides an example of the frame object, which includes a subtype object representing the \texttt{PING} frame and a generic action \texttt{handle(f:frame)} that must be implemented by subtype objects. The \texttt{PING} frame defines a \texttt{data} field containing the four-byte payload as described in the specification.

\begin{codeInput}{c}{State example}{state} 
object frame = {
  type this
  object ping = {
    variant this of frame = struct {  
        data : stream_data  
    }
  }
  action handle(f:this) = {  
    require false; 
  }
}
\end{codeInput} 

Ivy's "action" statement is used to manipulate the states and add requirements. An action can be considered as a procedure and cannot be stored in variables or passed as arguments. Listing \ref{inputPrg:action1} illustrates the \texttt{handle(f:frame)} action, which encompasses all the properties associated with the \texttt{PING} frame and adds requirements that will be checked every time a \texttt{PING} stream is received and generated. Lines 7 and 8 specify that the data payload must be a "ping" and have a length of four bytes. Line 9 requires that a \texttt{PING} frame should not be present in a packet using the relation \texttt{ping\_frame\_pending} defined on Line 1. Line 12 illustrates the invocation of the \texttt{enqueue\_frame(f:frame)} action, which also modifies various states within the model and is used to append a frame to a packet object.

\begin{codeInput}{c}{Object procedure example}{action1} 
relation ping_frame_pending

object frame = {
  object ping = {
    action handle(f:frame.ping)
    around handle {
      require f.data = ping_data;  
      require f.data.end = 4;
      require $\sim$ping_frame_pending;
      ...
      ping_frame_pending := true;
      call enqueue_frame(f);
    }
  }
}
\end{codeInput}

\begin{figure}[h!]
  \centering
  \includegraphics[width=1\columnwidth]{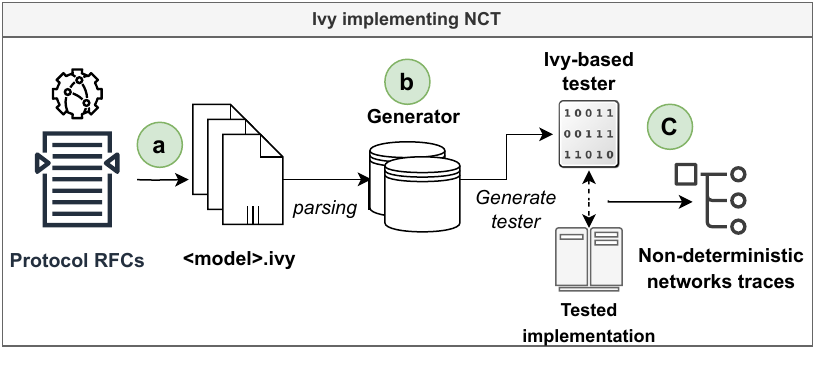}
  \caption{Ivy implementing NCT}
  \label{fig:fsm}
\end{figure}

\paragraph{\textbf{Network-centric Compositional Testing methodology (NCT)}}
NCT, a specialized approach within Model-Based Testing (MBT), is specifically designed for network protocols. It provides a structured method for creating formal specifications of Internet protocols and subsequently testing them \cite{McMillan_Zuck_2019}. The NCT principle is demonstrated in Figure~\ref{fig:fsm} using Ivy. The process begins with converting the RFC into an Ivy formal model \textcircled{a}. Once the Ivy code is parsed, a generator is used to create concrete and randomized testers \textcircled{b}. Finally, the implementation of the real-life protocol is tested and verified against the testers that employ an SMT solver to satisfy the constraints of the formal protocol requirements. When a requirement fails, the resulting traces \textcircled{c} can be analyzed to identify any potential errors or vulnerabilities. 

The NCT principle is based on compositional testing, which views formal specifications as a set of interconnected components/processes with their corresponding inputs and outputs. This approach allows testing protocol behaviors as observed on the wire, rather than relying on an abstract mathematical model of the protocol. This is why this methodology is called "network-centric".

\begin{figure}[h!]
  \centering
  \includegraphics[width=1\columnwidth]{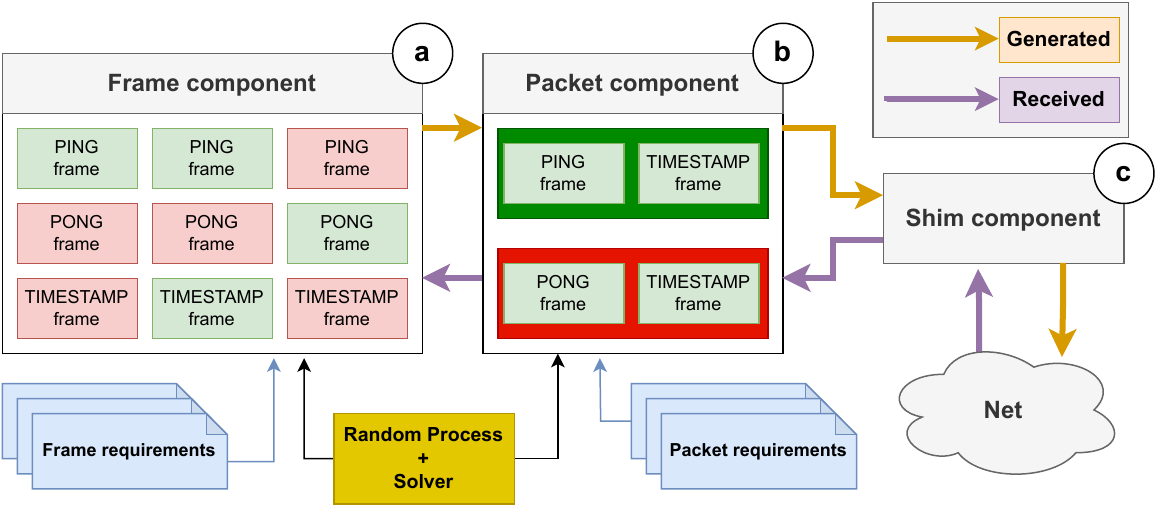}
  \caption{MiniP Network-Centric Testing (NCT) structure}
  \label{fig:ncm}
\end{figure}

\paragraph{} The design of \texttt{MiniP} is in line with the principles of the NCT methodology. In \texttt{MiniP}, the "Frame" process produces output that serves as input for the "Packet" process. The "assumptions" regarding the inputs of a process are treated as "guarantees" for the outputs of other processes. Figure~\ref{fig:ncm} provides a visual representation of this structure. In the context of \texttt{MiniP}, each element represents a layer of the \texttt{MiniP} stack, including the frame layer \textcircled{a} and the packet layer \textcircled{b}. The \textit{shim} component \textcircled{c} is responsible for transmitting and receiving packets across the network. When a packet is received, the \textit{shim} invokes the \texttt{ping\_packet\_event} action. This action contains all the specifications associated with the \texttt{MiniP} packet and will generate an error if any of the requirements are not met. For instance, it verifies that a packet always contains two frames in the correct order. The frames are similarly managed with their respective actions. In Figure \ref{fig:ncm}, the set of requirements is connected to the packet component \textcircled{b}.

\paragraph{\textbf{Limitations}} NCT's success derives from how effectively it identifies errors and vulnerabilities in real-world protocols like \texttt{QUIC} \cite{Crochet_Rousseaux_Piraux_Sambon_Legay_2021,McMillan_Zuck_2019a}. However, NCT presents some limitations associated with its inability to test time-varying network properties. 

For instance, it cannot model the congestion control mechanism specified in \texttt{RFC9002}. Calculating the time needed for packet generation and verification of received packets can be significant, particularly when implementations under test generate packets in bursts, leading to false congestion due to increased round-trip time. 

Additionally, debugging implementations is challenging because traces have no guarantee of reproducibility. In summary, here are the current main limitations of NCT for protocol testing:

\paragraph{{\normalfont\textcircled{1}} Lack of the expressiveness to handle time-varying network properties \cite{McMillan_Zuck_2019a}.} NCT lacks the necessary capabilities to reason about precise time intervals and deadlines. It also does not provide guarantees on thread-scheduling or computation time. For example, NCT cannot verify whether a \texttt{MiniP} server will respond to a \texttt{PING} within three seconds.

\paragraph{{\normalfont\textcircled{2}} Non-reproducible experiments.} While NCT offers deterministic verification for formal properties, this determinism does not extend to the network (\textit{network nondeterminism}) or the implementations being tested (\textit{internal nondeterminism}). As a result, the experiments cannot be reproduced. For example, a \texttt{MiniP} server that crashes when sending an odd \texttt{TIMESTAMP} will not crash deterministically when tested with NCT. 

\paragraph{{\normalfont\textcircled{3}} Computational time exceeding protocol timeout.} The computation time required to verify incoming packets and generate packets that satisfy the model can interfere with the standard behavior defined by some protocols. This also hinders the reproducibility of the experiment. For instance, a \texttt{MiniP} server implementation in NCT may take too long to check the integrity of a \texttt{PING} and exceed the response time limit of 3 seconds. This issue would not occur with \texttt{MiniP} implementation on a modern computer, but it arises with real protocols due to the inherent complexity of their RFCs, as with \texttt{QUIC} \cite{McMillan_Zuck_2019a}.

\newpage
\section{Network Simulator-centric Compositional Testing}\label{sec:approach}

Network Simulator-Centric Compositional Testing (NSCT) is a specialized approach within Model-Based Testing (MBT) that aims to overcome the limitations of NCT discussed previously. NSCT, similar to NCT, adopts a network-centric perspective and employs a combination of two key ingredients.

\paragraph{\textbf{Ingredient 1: Introduction of Network Simulators (NS)}}

\paragraph{\textbf{Type of Network Simulators}} NS tools permit running a model or a real executable inside a controlled network environment. Model-oriented simulators are mainly used to verify protocols during their development stage \cite{breslau2000advances}. Many types of NS exist; we will focus on time-dependent NS tools that have two main properties: they proceed chronologically and maintain a simulation clock \cite{issariyakul2009introduction}.
This clock is essential to verify time-related properties. There are two types of \textit{time-dependent }NS: time-driven and event-driven. 

\paragraph{(1) Time-driven NS}  advance their clocks exactly by a ﬁxed interval of $\delta$ time units \cite{issariyakul2009introduction}.
This means that the simulation has a time precision of $\delta$.
To increase precision, $\delta$ must be small, which slows down the simulation computation.

\paragraph{(2) Event-driven NS,} by comparison, advance their clocks by variable steps. Such tools progress as events unfold; each time step jumps precisely to the next event:

\begin{algorithm}
\NoCaptionOfAlgo
\caption{Event-driven network simulator \cite{fujimoto2001parallel}}
\While{simulation is in progress}{
remove smallest time-stamped event; \\
set time to this timestamp; \\
execute event handler;}
\end{algorithm}

\paragraph{\textbf{Our approach using NS}} Our solution involves integrating NCT with an event-driven, time-dependent network simulator to effectively overcome the limitations \textcircled{2} (\textit{Non-reproducible experiments}) and \textcircled{3} (\textit{Computational time}), and partially limitation \textcircled{1} (\textit{Time-varying network properties}).

\paragraph{Addressing Limitation} \textcircled{2}: Model-based testers and protocol implementations are real software components that must be executed rather than just modeled. When they are executed in a controlled environment, it becomes possible to stabilize and replicate desired random behaviors like encryption. This solution addresses the second limitation \textcircled{2}. This ensures the determinism of model-based protocol testing.

\paragraph{Addressing Limitation} \textcircled{3}: Formal specifications developed to test internet protocols with respect to NCT focus on packet events. This means that the latency of the (network) link between the model-based test and the implementation under test (IUT) determines the clock steps.  Assuming a latency of $l$, if the IUT sends a packet at time $t$, the model-based tester will receive it at time $t+l$. If the model-based tester responds immediately (without waiting for a specific delay to verify a timing property), the IUT will receive the response at time $t+2l$. This resolves \textcircled{3} as the computation time does not affect the time perceived by the IUT.

\paragraph{Addressing (partially) Limitation} \textcircled{1}:
To address \textcircled{1} (\textit{time-varying network properties}), it is essential to employ a time-dependent NS. However, simply using a time-dependent NS is not enough. It is also necessary to have a formal verification tool that can interact with the simulation clock. We will discuss this further in the second ingredient of NSCT.

\paragraph{Additional values using NS:}
The use of NS provides the ability to manage and control the network. It simplifies the creation of various network-related situations, including connection migration as described in \texttt{RFC9000}. An NS also enables realistic simulation scenarios of advanced modern network protocols~\cite{Fujimoto2007}. In the following sections, we will introduce the specific NS that we used for NSCT.

\paragraph{\textbf{Specific NS}}
There are two modern discrete-event network simulators: 

\paragraph{(1) The ns-3\cite{riley2010ns}} simulator is a freely available tool that has been specifically designed for research and educational purposes. It operates using models. To enable the execution of direct code within \textit{ns-3}, the \textit{DCE} framework \cite{tazaki2013direct} intercepts system calls and links them to \textit{ns-3}. However, \textit{DCE} does not support many of the necessary system calls required for protocol implementation, and the environment it supports has become outdated. The use of \textit{ns-3 DCE} to simulate \texttt{QUIC} implementations requires significant effort and inhibits tool longevity.  Researchers have recently expressed concerns about various challenges encountered while trying to simulate \texttt{QUIC} implementations using the \textit{ns-3 DCE} framework \cite{Google_2023}.

\paragraph{(2) Shadow \cite{jansen2011shadow, 280766}} is a free, open source simulator. It was primarily designed to simulate Tor networks. Shadow works by intercepting a subset of the system calls (syscalls), simulating network calls. Despite lacking support for some key system calls initially, the Shadow project remained highly active and has since added built-in implementation for several important syscalls that were originally missing. This is a positive sign for the long-term sustainability of the tool.

Shadow provides a range of network-specific functionalities that are highly beneficial for researchers and engineers. It allows users to carefully design the network topology that they want to simulate, specifying nodes, links, and their connections. Shadow allows users to adjust parameters like link latency and jitter, which are crucial for evaluating the performance of networked applications under different network conditions. Beyond its flexible design capabilities (e.g., configurable topologies), Shadow offers live debugging features to monitor and troubleshoot network behaviors in real-time during simulations. 

Shadow easily supports single-threaded and multithreaded implementations.  For example, if there is a multithreaded server connected to two clients, Shadow enables deterministic debugging of one client while allowing the other client and the server to operate independently within the simulation. This level of control and precision in debugging, even in complex multithreaded scenarios, provides researchers and developers with valuable insights into the behavior and interactions of various components. As a result, it improves the thoroughness of protocol analysis.

\paragraph{\textbf{Ingredient 2: Integration of time-varying network properties testing in Ivy}} 

\paragraph{}In practice, network link properties are designed in the Ivy model or directly with Shadow. Ivy then builds the simulation configuration file with those properties and references the executable used for the test. Finally, Shadow launches the IUT and the Ivy test.

\paragraph{Adapting Ivy for Event-Driven Network Simulation} 
Our approach aims to enhance the compatibility of the Ivy verifier with event-driven network simulators. To improve protocol verification in Ivy, we propose an adaptation that introduces an interface for manipulating time-related actions/relations. The interface is implemented in C++ and leverages the \textit{'time.h'} library to facilitate the interception of system calls (syscalls) by the Shadow simulator.

This interface provides several key functionalities, including the manipulation of time in various units (seconds, milliseconds, microseconds), timer control (start and stop actions), and current-time querying. The interface supports setting time breakpoints at specific events and implements both blocking and non-blocking sleep mechanisms. Using non-blocking sleep allows the simulation to receive network events while "sleeping," resulting in simulations that more accurately mirror real-world network behavior. 

\begin{figure}[t!]
  \centering
  \includegraphics[width=1\columnwidth]{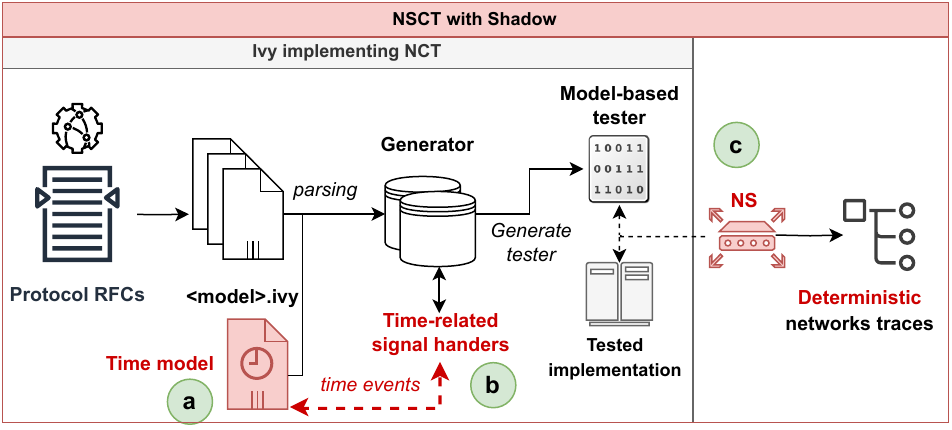}
  \caption{Protocol Formal Verification toolchain (PFV)}
  \label{fig:fsm2}
\end{figure}

Ivy's time interface is extensible, which allows for further adaptations and enhancements to meet the evolving needs of network protocol verification. The time interface is represented at Figure \ref{fig:fsm2} \textcircled{a}.

Our progress in Ivy for simulating event-driven networks is built on an improved method for controlling event generation. We use signal handlers along with time-based signals such as \texttt{SIGALRM} to accurately manage event timing. This approach is illustrated in Figure \ref{fig:fsm2} \textcircled{b}. It is especially effective in situations that require delayed responses, as it allows precise control over the timing of event generation and processing.

Finally, while Shadow's ability to modify network conditions is beneficial, it lacks flexibility, as it cannot vary the delay during the connection. To address this limitation, we developed a formal model that represents network quality, enabling us to simulate more specific scenarios of network condition. Nevertheless, we still rely on Shadow for reproducibility and intercepting time syscalls as shown in Figure \ref{fig:fsm2} \textcircled{c}.

\paragraph{Monitoring time-varying properties} We can now use Shadow intercepting time syscalls to define safety properties for time-varying properties in Ivy without having to modify the tools directly. Using the implemented time module and the standard Ivy key words such as "\texttt{require}" or "\texttt{assume}", we can model all the first-order logic formula with time as variable or predicate. 

\paragraph{Protocol Formal Verification (PFV)} PFV\footnote{\url{https://github.com/ElNiak/PFV}} toolchain implements NSCT by leveraging Ivy and Shadow. The usability of the previous work is enhanced by implementing a multistage docker containerisation procedure  \cite{merkel2014docker}, coupled with microservices and a basic graphical interface to initiate experiments. This architecture allows for easy testing of new protocol implementations with Ivy and Shadow. All containers implement a REST API to start Ivy.

\paragraph{\textbf{Case Study - \texttt{MiniP} Protocol}}  

In our \texttt{MiniP} formal specification example, we implemented the property that the \texttt{PONG} message should be received within 3 seconds after the \texttt{PING} message being sent, as seen in Listing \ref{inputPrg:pong}. To achieve that, we use the concept of time breakpoint. Then we add requirement manipulating the values extracted from these breakpoints.

\begin{codeInput}{c}{Testing time-varying properties}{pong} 
# Get current time from last break point 
current_time := time_api.c_timer.now_millis_last_bp; 
# Check that it satisfies the 3 seconds limit   
require current_time $\leq$ 3000;       
\end{codeInput} 

The time breakpoint is set in the  \texttt{PING} frame event handler as presented in Listing \ref{inputPrg:ping}:
\begin{codeInput}{c}{Adding time breakpoint}{ping} 
object ping = {
    around handle {
      # [previous requirements]
      ...
      call time_api.c_timer.start; #add time break point
      # [previous requirements]
    }
}
\end{codeInput}

This example demonstrates how simple it is to test a time-varying network property with a safety property thanks to the network-simulator assumption. 

Three distinct implementations of the protocol were discerned.
The first implementation consistently adhered to the specification by responding with \texttt{PONG} within the 3-second limit.
In contrast, the second implementation displayed intermittent deviations from the desired behavior, indicating the necessity for further refinement.
The third implementation consistently failed to meet the specification, exposing significant deficiencies.

Shadow's capabilities were employed to introduce link jitter between the client and the server, simulating network conditions with varying packet delivery times.
This additional element of uncertainty influenced the performance of the implementations.
The previously flaky implementation, which occasionally deviated from the specification, now violated the time constraint more frequently.

The experiment's determinism helped us to identify a specific seed value that leads to early connection failure in the flaky implementation.

Shadow's debugging capabilities allow precise analysis of the two faulty implementations.
By attaching a debugger to the implementations during testing, we can precisely identify which components were responsible for the deviations from the specified behavior.

\section{Threat to validity for \texttt{QUIC}}\label{sec:quicverif}

This section provides an overview of how NSCT is applied to the \texttt{QUIC} protocol. It begins by defining the \texttt{QUIC} protocol and then discusses the modifications made to the formal model described in previous work \cite{Crochet_Rousseaux_Piraux_Sambon_Legay_2021,McMillan_Zuck_2019a}. It also includes an analysis of the results obtained from the \textit{picoquic} implementation, highlighting a specification violation. This issue was subsequently addressed through a pull request in the \textit{picoquic} repository, fixing the error and aligning the implementation with the \texttt{QUIC} specifications.

\paragraph{\texttt{\textbf{\texttt{QUIC}}}} is a modern transport protocol that combines the advantages of \texttt{TCP} (Transport Control Protocol) and \texttt{TLS} 1.3 (Transport Layer Security), while overcoming their limitations, as detailed in \texttt{RFC9000} \cite{9000}. It introduces innovative secure communication methods at the transport layer. The RFC describes how data are organized into frames and packets to ensure effective data segmentation, reliability, and control.

\paragraph{\textbf{Tested implementation:}} \textit{picoquic} is a research implementation of \texttt{QUIC} \cite{picoquic}. This implementation participated in the development of a \texttt{QUIC} standard by providing feedback. \textit{picoquic} is written in C and consists of 103k lines of code. The tool incorporates various \texttt{QUIC} extensions and is currently under active development, making it an ideal choice for testing purposes. Moreover, \texttt{QUIC} has recently been chosen as the basis of \texttt{HTTP/3} and is expected to handle a substantial portion of internet traffic in the coming years~\cite{McMillan_Zuck_2019a}.

Table \ref{tab:summary} summarizes the contribution to \texttt{QUIC} formal specification and the problems we found per RFC while testing \textit{picoquic}:

\begin{table}
\centering
\begin{tabular}{|l|l|l|l|} 
\hline
            & \textbf{A. \texttt{RFC9000}}                          & \textbf{B. \texttt{RFC9002}}                                        & \textbf{C. Ack Frequency}                         \\ 
\hline
\textbf{Previous works} & \begin{tabular}[c]{@{}l@{}}Partially \\ complete\end{tabular}    & \multicolumn{1}{c|}{/}                                       & \multicolumn{1}{c|}{/}                          \\ 
\hline
\textbf{Contributions}  & \begin{tabular}[c]{@{}l@{}}- Ack-delay\\ - Idle timeout\end{tabular} & \begin{tabular}[c]{@{}l@{}}- Congestion control\\ (rtt calculation)\\ - Loss recovery\end{tabular} & 90\% of the draft                             \\ 
\hline
\textbf{Problems found} & \begin{tabular}[c]{@{}l@{}}Max \\ retransmission\end{tabular}    & \multicolumn{1}{c|}{/}                                       & \begin{tabular}[c]{@{}l@{}}Misinterpretation in\\ a frame field\end{tabular}  \\
\hline
\end{tabular}
\vspace{0.4cm}
\caption{Summary of contributions to Ivy model and problems found in \textit{picoquic}}
\label{tab:summary}
\end{table}

\paragraph{\textbf{A. Analysis of \texttt{RFC9000}:}} Our approach, integrating the time module and Shadow, has enabled enhancements to the existing \texttt{QUIC} model by incorporating time-related requirements as per \texttt{RFC9000} specifications.

We focused on the idle timeout connection termination behavior of \texttt{QUIC}. \texttt{QUIC} outlines three primary methods for connection termination: immediate close, stateless reset, and idle timeout. We designed our tests to validate the implementation of these methods, particularly the idle timeout.

According to \texttt{RFC9000} section 10.1, an endpoint restarts its idle timer upon receiving or sending ack-eliciting packets (i.e., the packet triggering ACK mechanism), ensuring that connections remain open during active communication. 

To prevent overly brief idle timeouts, \texttt{QUIC} mandates an idle timeout period be at least three times the current Probe Timeout (PTO). This extension allows multiple opportunities for packet transmission before a timeout. 

The connection in \texttt{QUIC} is automatically and \textit{silently} closed, discarding its states if it remains idle beyond the minimum duration set by the \texttt{max\_idle\_timeo\\ut} transport parameter.  

Our experiments revealed some discrepancies in the implementation of the idle timeout feature not in line with the standard behavior dictated by \texttt{RFC9000}. We noticed deviations in the handling of retransmission thresholds and idle timeouts.

This test involved suspending packet transmission after a random period and observing if the connection closes silently, according to the specifications. However, our experiments revealed a deviation in the picoquic implementation. Rather than closing the connection after the idle timer expired, picoquic terminated it prematurely upon reaching a retransmission threshold. This behavior, probably influenced by \texttt{TCP} retransmission mechanism, deviates from \texttt{RFC9000} standards, which require explicit notification through \texttt{CONNECTION\_CLOSE} or \texttt{APPLICATION\_CLOSE} frames for such terminations.

The discovered issue has been resolved through a pull request that was merged into the picoquic repository. This confirms the effectiveness of NSCT in detecting real-world anomalies in protocols.

\paragraph{\textbf{B. Analysis of \texttt{RFC9002}} \cite{9001}} This RFC discusses loss detection and congestion control in \texttt{QUIC}, differentiating it from \texttt{TCP}. It includes 37 mandatory specifications and 27 recommendations as per \texttt{RFC2119}. Key concepts introduced include the "probe timeout" (PTO) for managing congestion windows and the round-trip time (RTT) estimation process, comprising metrics like "min\_rtt", "smoothed\_rtt", and "rttvar". The RFC also details a sender-side congestion control mechanism, akin to \texttt{\texttt{TCP}/NewReno}, focusing on packet losses and Explicit Congestion Notification (ECN) \cite{Floyd_Ramakrishnan_Black_2001,Black_2018} .

Our analysis involved implementing and testing the specified requirements and behaviors of \texttt{RFC9002} in the context of congestion control and loss recovery, excluding the ECN component because it requires kernel support, which is not currently supported by Shadow. Future work could explore additional congestion control mechanisms like \texttt{CUBIC} \cite{Rhee_Xu_Ha_Zimmermann_Eggert_Scheffenegger_2018a} or \texttt{BBR} \cite{Cardwell_Cheng_Yeganeh_Swett_Jacobson}, and extensions such as \texttt{\texttt{QUIC}-FEC} \cite{8816838}. Tests were conducted to evaluate the model's behavior under various network conditions, including loss, delay, and jitter, ensuring adherence to the RFC's guidelines.

While our formal specification of the \texttt{RFC9002} did not identify specific problems in the \textit{picoquic} implementation, it significantly contributed to refining the formal specification of \texttt{QUIC}, making it more precise and closely aligned with real-world scenarios.

\paragraph{\textbf{C. Analysis of "\texttt{QUIC} Acknowledgement Frequency" \cite{Iyengar_Swett_Kuhlewind} extension}}   
Currently in its \texttt{draft-05} version, this extension enhances \texttt{QUIC} by allowing for delayed packet acknowledgments. It introduces the \texttt{min\_ack\_delay} transport parameter and two new frames: \texttt{ACK\_FREQUENCY} and \texttt{IMMEDIATE\_ACK}. The \texttt{ACK\_FRE\\QUENCY} frame adjusts acknowledgment rates based on the network state, while the \texttt{IMMEDIATE\_ACK} frame assists connection liveliness.

In our examination, we analyzed the integration of this extension into the \texttt{QUIC} formal specification, focusing on the implications of delayed acknowledgments. During this process, an error was identified in the picoquic implementation; it incorrectly returned a \texttt{FRAME\_ENCODING\_ERROR} when processing an \texttt{ACK\_FREQUENCY} frame. This issue, initially suspected to be a draft inconsistency, was actually due to a misinterpretation of the "ACK-Eliciting Threshold" field in \textit{picoquic}.

\section{Discussion and future work}\label{sec:futur}
\paragraph{ }It is clear that the NSCT methodology, which involves various tools, is successful in uncovering new behaviors in protocol implementations, especially in terms of their temporal dynamics. However, combining multiple tools also brings about the intrinsic limitations of each tool and difficulties from their joint use. For example, employing network simulators, such as Shadow, comes with specific constraints. The necessity for simulators to depend on system calls for interacting with implementations restricts the range of implementations that can be tested. Moreover, busy loops (an anti-pattern not very used) reduce precision in time as they do not wait through system calls. In addition, the topology part of the NSCT testing process requires scenario-based simulations, where the behavior of the network within the simulated environment is predetermined. 

A natural strategy to tackle these limitations is to replace certain tools used in the existing NSCT framework. For example, employing a network simulator that accommodates dynamic topology might alleviate the restrictions of scenario-based simulations, though it would necessitate modifications to existing automation scripts. Furthermore, substituting Ivy with any tool that implements the NCT methodology could be feasible, but this would require adapting the Ivy models.

In addition to improvements related to the joint use of several tools. Other future avenues may also be investigated. For example, a further improvement of the time module can allow for a more thorough verification of different types of properties. Closer integration between Ivy's generation process and the time module could make generating events at specific time points easier. This adaptation would enhance the tool's precision and overall usefulness. 

Expanding the scope of the testing to include different congestion mechanisms in various protocols would provide more detailed insight into the details related to implementation and effectiveness. 

Applying the methodology and tools discussed here to a broader range and scale of \texttt{QUIC} implementations would better validate and improve reliability and security.

Another promising area of research lies in the examination of the synergy between formal attacks models in network and Shadow's capabilities, especially considering the ongoing development of protocols like \texttt{MPQUIC} \cite{DeConinck2017}. Currently, \texttt{MPQUIC} is still in its draft phase, grappling with significant security considerations between two main solutions. Given Shadow's unique ability to create custom network topologies, our methodology stands to offer substantial assistance. It enables the modeling of both solutions under consideration for \texttt{MPQUIC}, providing a comprehensive framework to assess their security implications and vulnerabilities.

Furthermore, another innovative approach is to leverage AI to simplify the creation of formal models from RFCs. This integration would greatly streamline the modeling process, making it more efficient and accessible. Additionally, a Graphical User Interface (GUI) for Ivy would allow users to engage in formal modeling without needing to understand the complexities of Ivy code, thereby making the process more user-friendly and approachable.

\section{Conclusion}\label{sec:conclo}

Protocol validation methods are diverse. Notably, the \textit{INET} suite within the \textit{OMNeT++}  simulation library offers a powerful tool for network protocol validation \cite{varga2010omnet++}. Previous research used \textit{INET} to simulate a \texttt{QUIC} model within \textit{OMNeT++} \cite{9142723,9524947}, but the evaluation was limited to protocol models and did not include real-world implementations. Studies like \cite{Bernardeschi2021} used formal verification methods and simulations to validate complex protocol properties. This work measured the impact of specific attacks like Denial-of-Service on network parameters like energy consumption and computational effort, but did not use formal methods alongside simulations for verification.

Other research \cite{leonard1997introduction, bolognesi1994converging, bolognesi1994timed} attempted to enable the expression of time-varying network properties in the ISO standard to specify OSI protocols, known as Language Of Temporal Ordering Specification (\textit{LOTOS}) \cite{bolognesi1987introduction}. While extensions to \textit{LOTOS} offered varied expressiveness \cite{bolognesi1994converging}, subsequent Model-Based Testing (MBT) tools like \textit{TorXakis}~\cite{tretmans2019model} only ensure guarantees about the tester's side due to its lack of a network simulator. This limits its ability to verify time-varying network properties. 

In \cite{li2021model}, the authors compose their model with a network model to control the non-determinism of the network. This approach increases determinism, but it is not as powerful as NSCT, which extends determinism to IUT. \cite{bishop2018engineering} proposed using a test oracle on IUT traces, which allows offline verification of time-sensitive network properties that change based on network conditions. This approach avoids the high computational cost associated with online verification. However, this approach does not allow for the expression of time-dependent scenarios. Additionally, verifying traces does not facilitate the reproducibility of errors.

In this study, we propose an extension to Network-centric Compositional Testing (NCT). NCT is a simulation-based formal verification approach previously employed to validate \texttt{QUIC} implementations with Ivy. However, it has certain drawbacks; Ivy and NCT cannot capture time-varying network requirements or replicate experiments due to the inherent randomness of the methodology and the network. In addition, the extensive computational time required to scrutinize actual implementations of Internet protocols may affect protocol behaviors.

This study has successfully demonstrated the efficacy of \textit{Network Simulator-centric Compositional Testing} (NSCT) in enhancing the verification of network protocols, particularly in addressing key challenges of NCT. NSCT, through the integration of the Ivy tool and the Shadow network simulator, effectively solves several issues. These include the addition of time-varying network property verification, ensuring deterministic outcomes in protocol testing, and enhancing the reproducibility of test results. Our paper demonstrates the described method using a custom minimalist \texttt{MiniP} protocol. The application of NSCT in the picoquic implementation of \texttt{QUIC} identified a compliance error with time-varying network specifications, which was then rectified. This underscores the methodology's capability in managing complex, real-world network scenarios. 

Additionally, a formal model is developed for \texttt{RFC9002} that integrates congestion control and loss recovery into the existing \texttt{QUIC} model. The formal model of the "Acknowledgement Frequency" \texttt{QUIC} extension is also included.

\paragraph{\textbf{Acknowledgement}} We would like to thank Maxime Piraux for his help to validate \texttt{QUIC} experiment results.

\paragraph{\textbf{Artefacts}} The artefacts of this paper are available at \url{https://zenodo.org/doi/10.5281/zenodo.10819552}.

\bibliographystyle{splncs04}
\bibliography{refs}

\end{document}